\documentclass{vgtc}                          

\graphicspath{{figures/}{pictures/}{images/}{./}} 

\usepackage{times}                     

\usepackage{tabu}                      
\usepackage{booktabs}                  
\usepackage{lipsum}                    
\usepackage{mwe}                       

\usepackage{mathptmx}                  

\usepackage{enumitem}
\usepackage{comment}

\onlineid{8211}

\vgtccategory{Research}

\vgtcinsertpkg

\title{Designing a Virtual Reality Training Apprenticeship for Cold Spray Advanced Manufacturing}

\author{Mahsa Nasri
\and Uttkarsh Narayan 
\and Mustafa Feyyaz Sonbudak
\and Aubrey Simonson 
\and Maria Chiu 
\and Jason Donati 
\and Mark Sivak 
\and Mehmet Kosa 
\and Casper Harteveld 
}
\affiliation{\scriptsize Northeastern University\thanks{e-mail:\{nasri.m, narayan.u, sonbudak.m, simonson.au, chiu.chin, j.donati, m.sivak, m.kosa, c.harteveld\}@northeastern.edu}}

\teaser{
  \centering
  \includegraphics[width=1\textwidth, height=7cm]{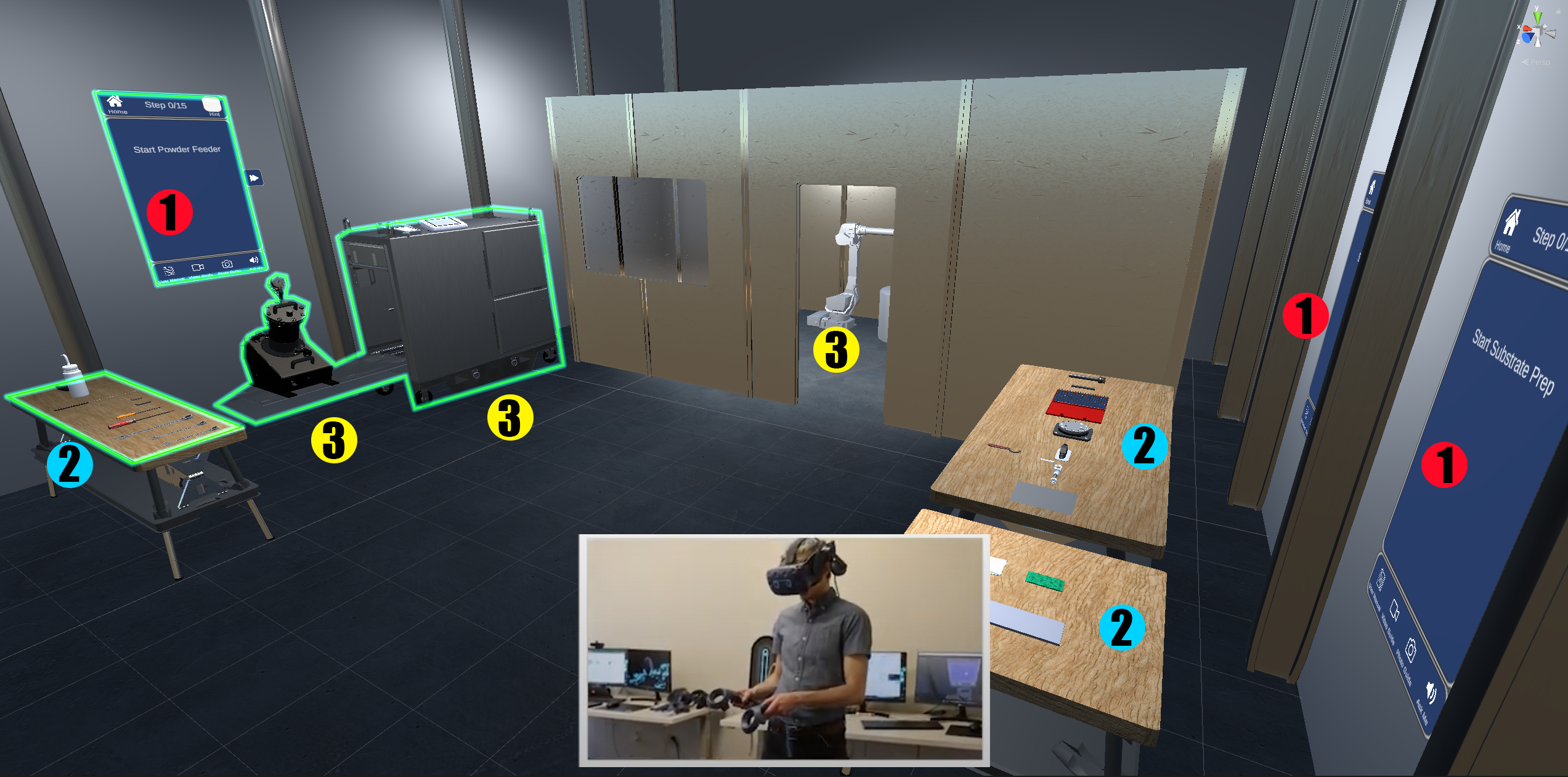}
  \caption{The VR apprenticeship environment: yellow circles (3) mark three of six modules of the Cold Spray procedure. The highlighted green area is the Powder Feeder module, where the participant is (dis)assembling. Red circles (1) represent task decomposition, and blue circles (2) indicate essential tools for each module. }
  \label{fig:teaser}
}

\abstract{%
Apprenticeship and training programs in advanced manufacturing frequently encounter safety and accessibility concerns due to using heavy machinery. Virtual Reality (VR) training addresses such constraints while maintaining the spatial and procedural learning requirements of such training. However, designing effective VR training is challenging because advanced manufacturing processes are complex and require experts to train novices for a long time. This paper presents a VR Training Apprenticeship (VRTA) tailored for cold spray, which we carefully designed to teach novices step-by-step this particular advanced manufacturing process. To assess its effectiveness, we conducted an exploratory study (\textit{n} = 22). We evaluated user experience (UX) measures in the form of quantitative scales, users' qualitative insights, and task performance with real-world machinery after the VR training. We discuss how the VRTA design contributed to the effectiveness and the challenges of considering VR training for advanced manufacturing. 
} 


\begin{document}


\firstsection{Introduction}

\maketitle

Technical and vocational education training in advanced manufacturing, such as cold spray, often involves hands-on apprenticeship programs, which can be risky, costly, and limited in availability \cite{snell_vocational_2019}. Virtual Reality (VR) training systems offer a safer, more accessible alternative for skill development in advanced manufacturing \cite{bailenson_experience_2018, choi_virtual_2015}.
VR training emphasizes active learning, promoting spatial knowledge acquisition and observation skills in real-world scenarios in a safe and hazard-free environment  \cite{bailey_assessment_2017}. Several VR training programs exist for advanced manufacturing, such as assembly training systems \cite{abidi_assessment_2019,choi_virtual_2015}, training for human-robot collaboration \cite{matsas_design_2017}, and simulating manual assembly for training purposes \cite{lopez_comparison_nodate, dwivedi_manual_2018}.
To provide cold spray engineers and technicians with a VR training alternative, we developed a Virtual Reality Training Apprenticeship (VRTA) that comprises all major sub-processes required to operate a cold spray system, carefully considering how we could teach such a complex advanced manufacturing process. We conducted an exploratory study to explore the effectiveness of our VRTA design. We focused on evaluating the assembly and disassembly of the powder feeder module, a primary and common task within the cold spray process. Our study confirms the VRTA effectiveness, with 22 participants completing the powder feeder module in VR and then the real-world task. We discuss design considerations that participants found helpful, along with identified challenges, offering insights for VR training in advanced manufacturing.

\section{VR Training Apprenticeship Design}
Cold spray 
is an advanced manufacturing technology that applies coatings of metallic or non-conductive substances to another surface through gas-powered high-velocity spray \cite{papyrin_cold_2006}. Safety is a real concern in the process, especially for novice operators. Hence, VR can provide valuable alternative training. Working collaboratively with experts, we dissected the cold spray process into six distinct modules (powder feeder, applicator, system leak check, substrate preparation, nozzle, and spray). 
Here, we outline the key considerations of our VRTA design, which, after several iterations, resulted in a training that takes approx. two hours to complete.\footnote{Videos and the VRTA itself are accessible here: \url{https://xert.co/}}

\textbf{Visual and Audio Realism}: Based on recent findings highlighting the influence of graphical realism in VR \cite{newman2022use}, and leveraging CAD models from cold spray equipment, we recreated the cold spray process on a 1-1 scale. 
We further simulated the audio in a typical cold spray environment (e.g., the sound of dust collectors) to improve the sense of immersion and presence. 

\begin{figure}
    \centering
        \centering
        \includegraphics[width=1.0 \linewidth]{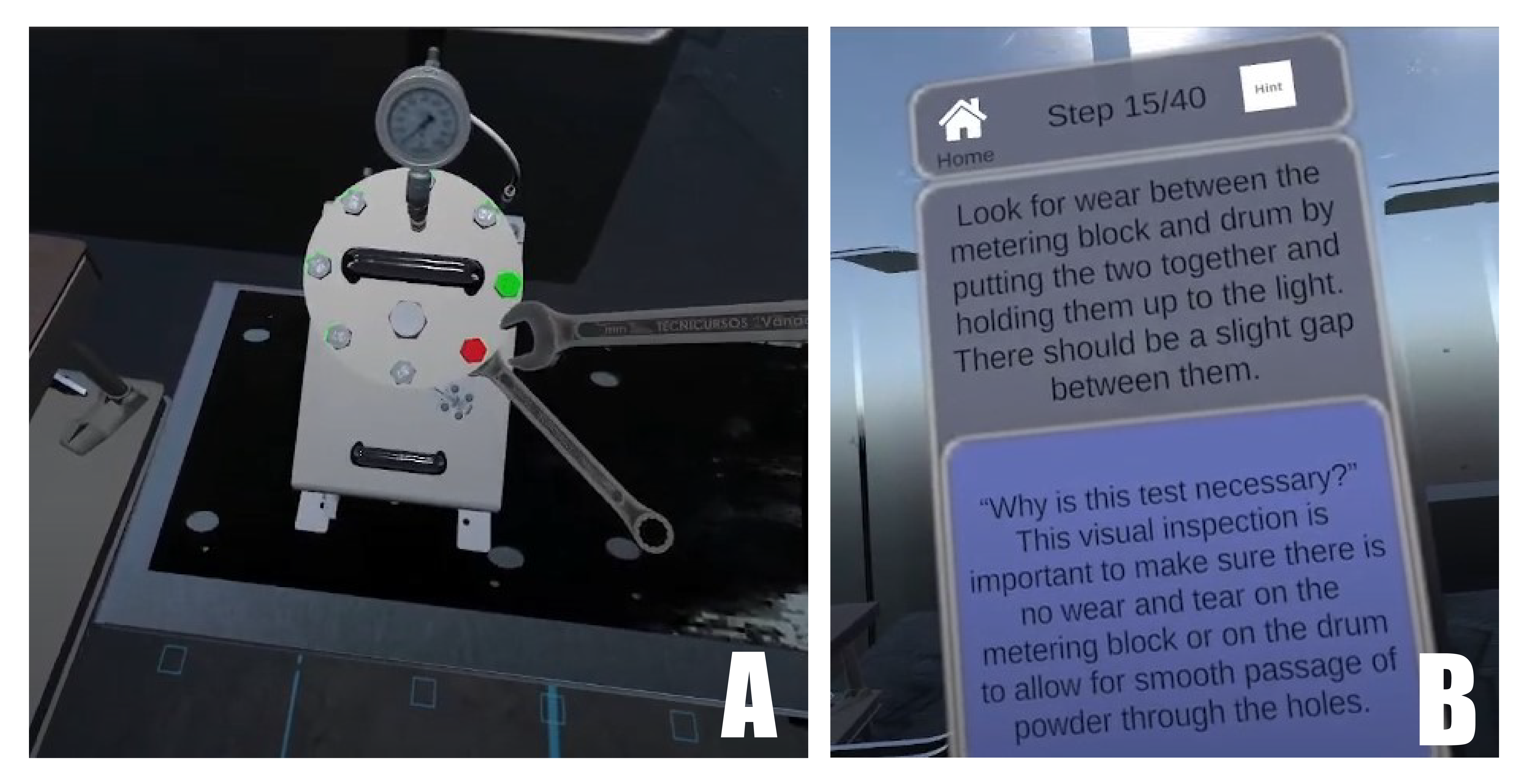}
        \caption{(A) Hint system: green and red highlight correct and incorrect interactions. (B) Task instruction: the gray box is task decomposition, and the blue box is the why statement.}
        \label{fig:whystatement}
\end{figure}

\textbf{Task Instructions}: Providing task instructions improves the cognitive engagement of participants compared to unguided training \cite{penalver2021providing}. 
For the task instructions in our VRTA, we first decomposed the tasks, inspired by the Four-Component Instructional Design (4C/ID) model \cite{van2017ten}, to optimize step size and break the process down based on a logical order and our learning goals. To help users understand the consequences of incorrect task completion, we included \textit{Why Statements} (Fig.~\ref{fig:whystatement}). Moreover, operators frequently need to convey information on names and functions. Hence, we consistently mentioned part names and their uses in relevant steps to help users become familiar with cold spray parts. Importantly, we placed the task instructions directly in front of the users so they could read the instructions while simultaneously engaging with the task itself (Fig.~\ref{fig:teaser}). 

\textbf{Hint System}:
Task instructions alone are insufficient \cite{ho_virtual_2018}. We created two hint systems: automatic and manual. Prior research indicates that users are reluctant to seek assistance and frequently ignore unclear hints \cite{aleven2000limitations}. To tackle this, automatic hints are activated 30 seconds into tasks, highlighting essential objects visually to guide users efficiently (Fig.~\ref{fig:whystatement}). Manual hints are available through the user interface to repeat text-to-speech audio for task instructions and highlight relevant objects.

\textbf{Task Interactions}: 
While we visually (and audibly) simulated the cold spray process on a 1-1 scale, we designed the actual \textit{Task Interactions} to highlight and focus on the important learning goals. For instance, the powder feeder (PF) assembly and disassembly steps involve 23 steps, 8 of which align with our learning goals. In these 8 steps, users must meticulously follow task instructions and engage with the system similarly to real-life scenarios. Minimal interaction is necessary for the remaining 15 steps as the VRTA auto-completes these steps. In addition, based on the literature \cite{brough2007towards}, to reinforce what has to be done in each step, we included \textit{Confirmation Animations}, which are animations that are triggered at the end of each step that demonstrate the most efficient method to complete the task. 

\section{User Study}
We conducted an exploratory user study to examine if the VRTA was effective. We had two evaluations: first, the VR user experience (UX), and second, the task performance after playing the VRTA with a real-world PF. This study only included assembly and disassembly of the PF module, which is the most complex. The study was conducted in person, and all participants used the same head-mounted display, Varjo VR-3, connected to a compatible desktop computer. Each participant's session was scheduled for 1.5 hours for both the VR and real-world tasks. 

After they had signed a written informed consent form, we showed an introductory video about cold spray to provide context to participants who did not have prior experience with the process. Then, the participants completed the VRTA experience. After completing the VRTA, the participants completed a questionnaire about their VR experience. Next, we asked them to run the assembly and disassembly on the real-world PF with the necessary tools, including Allen keys and screwdrivers, under expert observation to prevent potential damage and to record the number of errors. Finally, participants completed an open-response questionnaire comparing their VR experience to their experience with the real-world task. 
We recruited 22 participants from a university campus with varying levels of VR familiarity via poster and email ads. 19 self-identified as male and 3 as female, aged 19-28 ($M$ = 23.33, $SD$ = 3.12 ). Participants received \$40 after completing their session.

We measured the number of errors during the real-world task to evaluate task performance. Validated by experts, we identified 10 potential errors (e.g., actions that may damage the PF or the operator's safety) and then examined the participants' videos to see what errors they had made. On the questionnaire, we used the usability and presence sub-scales from the VRUSE instrument \cite{kalawsky_vrusecomputerised_1999} and the NASA Task Load Index (NASA-TLX) to assess cognitive load \cite{hart_nasa-task_2006}. We aggregated the Likert scale data for the usability, presence, and NASA-TLX scales according to their respective guidelines \cite{kalawsky_vrusecomputerised_1999,hart_nasa-task_2006}. For qualitative open responses, we used inductive open coding \cite{braun_using_2006} and consolidated the codes to generate insights into the participants' experience.

\section{Preliminary Results}
Overall, we observed that the VRTA is usable and does not cause excessive cognitive load; second, participants could perform the tasks on real-world PF. We also analyzed qualitative feedback 
to examine to what extent our design contributed to this success. 

\subsection{UX and Performance Measures}
The participants achieved Mean (\textit{M}) scores of 3.70 ($SD$ = 0.63) on the usability scale and 3.83 ($SD$ = 0.54) on the presence scale, both out of 5 points. The average score for the self-reported overall usability was 3.27 ($SD$ = 0.83), and for the presence was 4.18 ($SD$ = 0.66) out of 5 points. The NASA-TLX sub-factor reports are as follows: Mental demand ($M$ = 4.27, $SD$ = 1.45), Physical demand ($M$ = 3.45, $SD$ = 1.60), Temporal demand ($M$ = 4.05, $SD$ = 1.40), Effort  ($M$ = 4.27, $SD$ = 1.28), Frustration ($M$ = 3.60, $SD$ = 1.87), and Performance ($M$ = 2.95, $SD$ = 0.90). A score of 7 indicates a high task load, reflecting an unfavorable outcome. All scores are satisfactory and acceptable \cite{kalawsky_vrusecomputerised_1999, grier_how_2015}.


As for task performance, 8 participants completed the real-world PF task without errors, 7 had one, and 6 had two errors. Only one participant had more than 3 errors out of 10 potential errors. 

\subsection{Participant Insights}
To further explore participants' experiences, we analyzed their feedback regarding the design components they found helpful or hindering their learning and performance.
The VRTA's \textit{visual and audio realism} received the most positive mentions, with participants appreciating the accurate representation of the PF and its parts. Even after performing the tasks, participants still referred to realism as a helpful and satisfying feature: \textit{An accurate scale of the working apparatus and the device gave the idea of what actually a powder feeder looks like}--P23.
In contrast, some participants described unrealistic aspects of the VRTA, such as the physical attributes of the tools or interactions: \textit{the weight of the wrench and the amount of torque required to unscrew and screw back on were not felt which would be important}--P4.

Many participants also highlighted the ``clear \textit{[task] instructions}'' (P11) and how these aided in understanding the process. They also mentioned that seeing the instructions alongside performing the task helped them remember the process and create a ``mental image of how the action must be done'' (P14). 
However, some participants found the information a lot to consider: \textit{... I don't feel much confident working on an actual powder feeder because I was not able to grasp all the information...}--P23. 

The \textit{hint system} received positive feedback from most participants; auto and manual hints were mentioned as a helpful feature: \textit{Explaining the process and highlighting the parts which were to be used were very helpful}--P12. Some participants requested more hints for some steps: \textit{How to place the spanner, bolts could have been highlighted, places where vacuum had to be put in.}--P17.

We also found that the \textit{confirmation animation} helped participants to understand the instructions better: \textit{The [animation] videos showing how each action is done in real-time after you got the correct answer was a great way of checking yourself in my opinion}--P10. 
However, the \textit{task interaction} was perceived negatively overall. According to participants, some interactions, such as screwing, were generally not smooth and ''unintuitive'' (P11.) After performing tasks, some mentioned that screwing ``felt easier in real life'' (P4).

\section{Discussion}
We developed a virtual reality training apprenticeship (VRTA) for cold spray and observed successful UX and real-world task transfer results. As for the latter, while there are only 10 potential errors, the Powder Feeder (PF) assembly and disassembly tasks are complex, requiring many steps (i.e., 23 in total). Seeing that more than one-third completed this without any error at all and almost everyone but one participant with few errors is evidence that learning transferred from the VR to the actual task. Our exploration of participants' feedback revealed the strengths of why the VRTA may have been effective and the challenges that may have hindered it from being so, which we discuss below.

\subsection{Strengths of the VRTA}
The common practice in designing VR training is replicating the learning environment to enhance participants' spatial and procedural memory, drawing on theories that emphasize the importance of realistic environments for learning or creating the digital twin of the environment \cite{slater_depth_1994,slater_place_2009}. We also followed this practice and received positive feedback. Specifically, we recognized three particular design considerations that have contributed to the positive outcomes of our VRTA beyond ensuring a realistic experience. 

\textit{Visual Task Decomposition}: We implemented a \textit{visual} task decomposition containing instructions and hint buttons in front of the users for each step. This visual task decomposition encouraged participants to look at the instructions quickly and facilitated pressing the hint button. Despite consistently conveying information on names and functions, we observed that the technical terminology and instructions can still overload frustrated novice users, a challenge noted in advanced manufacturing training \cite{mcadam_absorbing_2014, coxhead_development_2019}. To mitigate this, we propose integrating a ``memory bank'' feature to archive tool names or embedding labels on parts, enhancing recall and learning efficiency.

\textit{Auto Hints}: Aligned with common practices \cite{ipsita_towards_2022}, our findings suggest that design features that added visual cues, such as highlighting parts in green for hints, were helpful for participants. Auto-hints, while helpful, were identified as insufficient for some learners. A possible solution is developing adaptive hint systems based on expertise to address diverse learning styles and needs. Nevertheless, this aspect underscores the need for future research to explore more effective ways to integrate hints within VR training.

\textit{Confirmation Animations}: Animations were triggered after completion of the step by users to show the correct orientation of tools and the correct way to perform the tasks. Participants reported that these animations were useful to reinforce what they had done or to correct potential misconceptions. The animations were thus a helpful tool for signaling correct actions and reinforcing learning through observation, which aligns with research on observational learning \cite{bandura_social_1977}.

\subsection{Challenges for the VRTA}
Previously, we suggested making the VR design less realistic by simplifying and abstracting certain steps in the VR. However, our findings show that certain aspects of the environment need to be \textit{more} realistic. Our participants specifically reported on actual touch or sensing and the interactions they perform in the VR instead of reality. 

\textit{Lack of Physicality}: Physical sensing of tools is crucial for developing muscle memory \cite{fourkas_kinesthetic_2008}. However, current VR technologies lack these multisensory experiences \cite{lawson_future_2016}. Initially, we did not anticipate that factors like the weight, torque, or pressure of tools would impact participants' experiences. Prior work also suggested \cite{herrington_immersive_2007} that physical realism is less important in learning than ``cognitive realism.'' However, our findings revealed that not sensing the physicality of tools in VR could frustrate some users. Some prior research proposes solutions to mimic weight using haptic feedback \cite{samad_pseudo-haptic_2019} or using gloves to mimic the pressure or torque \cite{jadhav_soft_2017,roberts_testing_2012}. However, wearing additional technology in VR training, especially for advanced manufacturing, could be challenging and disrupt learning. Thus, further research is needed, and we recommend balancing the importance of physicality and learning goals.

\textit{Non-Intuitive Interactions}: Considering VR's potential in advanced manufacturing training, we acknowledge the need for intuitive interfaces \cite{dangelmaier_virtual_2005} and the challenge of integrating current VR controllers in assembly and disassembly tasks. Generally, gestures in VR interactions often do not mirror real-life actions like tightening bolts or using a screwdriver. We observed during the actual PF disassembly that participants preferred to use their hands instead of tools whenever possible. Hence, one possible solution could be to use a VR headset with hand-tracking capability for specific interactions, such as removing the bolts. These challenges suggest that while VR has potential in advanced manufacturing, further development is needed to realize its full benefits and drawbacks.



\subsection{Limitations}
We evaluated only one module. Although the selected module is the most complex and all modules share the same design format, the generalizability of our findings to other modules remains an open question, and, thus, whether the complete two-hour VR training is effective. We also acknowledge that our population sample is small, gender-biased, and, due to its voluntary nature, potentially biased to technologically proficient or favorable individuals towards VR. Importantly, we worked with university students, and the target audience is future cold spray engineers and technicians who may respond differently to how information is provided. Aside from scaling and diversifying this work in terms of scope and participants, future work can include comparative studies with other training interventions or design variations of the VRTA to understand its effectiveness better. Above all, for generalizability to VR training in advanced manufacturing, a deeper understanding of what VR design considerations are critical for effective VR training requires theoretical advances in the field.   


\acknowledgments{
This material is based on work supported by the National Center for Manufacturing Sciences (NCMS). Any opinions, findings, or conclusions expressed in this material are those of the authors and do not reflect the views of NCMS. The authors further thank the Kostas Research Institute at Northeastern University, Naval Sea Systems Command, VRC Metal Systems, and all the cold spray experts, students, and developers who participated in the design of the VRTA.}

\bibliographystyle{abbrv-doi}

\bibliography{ref}
\end{document}